\documentclass[12pt]{amsart}

\usepackage{hyperref}
\usepackage{times}
\usepackage{amsfonts}
\usepackage{amsmath}
\usepackage[psamsfonts]{amssymb}
\usepackage{amstext}
\usepackage{amsthm}
\usepackage{latexsym}
\usepackage{bbm}
\usepackage{color}
\usepackage{graphicx}
\usepackage{subfig}
\usepackage{enumerate}
\usepackage{algorithm}
\usepackage{algorithmic}
\usepackage{url}
\usepackage{dsfont}
\usepackage{multicol}
\usepackage{subfig}
\usepackage{appendix}

\makeatletter
\newtheorem*{rep@theorem}{\rep@title}
\newcommand{\newreptheorem}[2]{%
\newenvironment{rep#1}[1]{%
 \def\rep@title{#2 \ref{##1}}%
 \begin{rep@theorem}}%
 {\end{rep@theorem}}}
\makeatother

\hypersetup{
  colorlinks   = true,
  urlcolor     = magenta,
  linkcolor    = red,
  citecolor   = blue
}

\newcommand{\mat}[1]{{\mathbf #1}}

\newcommand{\N}{\mat{N}}

\newcommand{\triplep}{(\Omega, \mathcal{F}, \mathbb{F} = (\mathcal{F}_t)_{(0 \leq t \leq T)}, \mathbb{P})}

\newtheorem{theorem}{Theorem}
\newreptheorem{theorem}{Theorem}

\newreptheorem{lemma}{Lemma}
\newtheorem{definition}[theorem]{Definition}

\newreptheorem{corollary}{Corollary}

\newreptheorem{proposition}{Proposition}
\newcommand{\expp}{ \mathbb{E}^{\mathbb{P}}}

\newcommand{\ignore}[1]{}

\title[Election Methodologies]
{Accurate Prediction of Electoral Outcomes}

\author{Dhruv Madeka}

\begin{document}

\begin{abstract}
    We present novel methods for predicting the outcome of large elections. Our first algorithm uses a diffusion process to model the time uncertainty inherent in polls taken with substantial calendar time left to the election. Our second model uses Online Learning along with a novel ex-ante scoring function to combine different forecasters along with our first model. We evaluate different density based scoring functions that can be used to better judge the efficacy of forecasters.
\end{abstract}

\maketitle

\section{Motivation}

Models of the presidential election are many and varied, each with it's own focus. There is a vast literature on methods to forecast the presidential elections. Models include those based on fundamental factors \cite{hummel2013fundamental}, Bayesian methods \cite{linzer2013dynamic}, and prediction markets \cite{berg2008results}. Other models including the popular FiveThirtyEight \cite{silver2016} combine multiple predictions through hybrid models which combine polls with other data. However, each of these methods suffers the same flaw, they donot incorporate the time uncertainty of the outcome. Using the same distribution at every point in time excludes the fact that a measurement made with greater calendar time to the election has more uncertainty than one made closer to the election (this is because, in filtration terms, a large amount of uncertainty can still be realized).

\begin{figure}[h]
\begin{center}
\includegraphics[width=16.5cm, height=7cm]{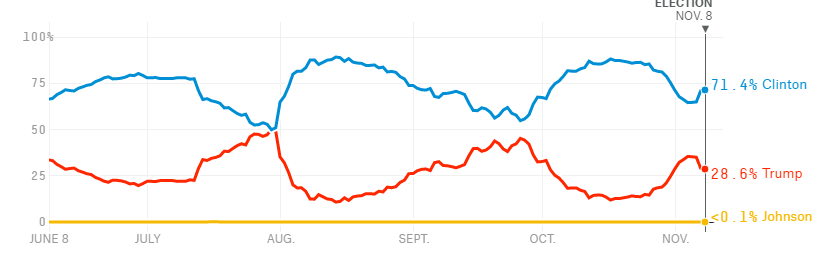}
\caption{FiveThirtyEight reported a 90\% probability in August, and a 55\% in October.}
\label{ns-ts}
\end{center}
\end{figure}

Consider Figure [\ref{ns-ts}], which shows the FiveThirtyEight probability time series. They report a 90\% probability in August, and a 55\% probability in October. If a forecaster truly believes that the probabilities move this much, then he should report 50\% \footnote{If we consider a Gaussian $\N(\mu, \sigma)$, and pick a level $\lambda$, as we increase the variance $\sigma^2$, the probability $\mathbb{P}(X \geq \lambda) \rightarrow 0.5$ as $\sigma^2 \rightarrow \infty$} (A rigorous proof and better elucidation of this can be seen in \cite{nassimelec}). The greater the calendar time to the Election (or Event) the greater time there is for uncertainty to be realized. The absence of this time component of uncertainty renders the model unstable. We propose a simple model, which utilizes a Brownian Motion with volatility. The Brownian Motion is a continous time stochastic process whose variance grows linearly with time. As a result, the greater the calendar time the more the uncertainty in the final realization. Inspired by the seminal CAPM Model in Finance, we propose a CAPM Model for the elections, treating each of the 50 states as a stock and the national popular vote as the market.

Second, we address the notion of comparing forecasters. The conventional method used, the Brier Score, fails to capture different aspects of forecasts that distinguish a skilled forecaster from an unskilled one. We propose the use of two methods to compare forecasters according to the density reported rather than just the probability. However, all of these comparisons are ex-post, namely they rely on the realization of the event. Ideally, we would like to be able to compare forecasters before the event so that we may trade the prediction markets on the event. We use an ex-ante method to compare forecasters in terms of a trading strategy, and then apply this combination to the US Election Betting Market. Our online mixture performs better than most experts.
\\ 

\section{Preliminaries} 
\label{sec:preliminaries}

\subsection{Brownian Motion}
\hfill\\

We work on a filtered probability space $\triplep$. We define a Brownian Motion $W(\omega)_t$ in the following way:

\begin{definition}
Any continuous time stochastic process $W(t): t \geq 0$ is a brownian motion if it has the following properties \cite{ks}:
\begin{itemize}
    \item For $t_1 \leq t_2 \leq ... \leq t_n 0$
    \begin{align*}
        W(t_n) - W(t_{n-1}), ..., W(t_2)-W(t_1), W(t_1) - W(t_0)
    \end{align*}
    are independent
    \item $W(t)-W(s)$ is distributed $\mathcal{N}(0, \sqrt{t-s})$ for $0 \leq s < t$; where $\mathcal{N}(\mu, \sigma)$ is the Normal Distribution with mean $\mu$ and variance $\sigma^2$
    \item the process $W(t): t \geq 0$ has almost surely continuous paths
\end{itemize}
\end{definition}

\hfill
\subsection{Proper Scoring Functions}
\hfill\\

Following \cite{savage1971}, consider $\triplep$, where $\mathbb{P}$ defines a convex class of probability measures on $\Omega$. A \textit{probabilistic forecast} is any probability measure $P \in \mathbb{P}$. A \textit{scoring function} is any extended real-valued function $S: \mathbb{P} \times \Omega \rightarrow \bar{R}$ such that $S(P, .)$ is integrable for all $\mathcal{P}$ in $\mathbb{P}$. We define the expected score under Q as:

\begin{align*}
    S(P, Q) = \int S(P, \omega) dQ(\omega)
\end{align*}

A scoring function is said to be strictly proper relative to $\mathcal{P}$ if:

\begin{align*}
    S(Q, Q) \geq S(P, Q)
\end{align*}

$\forall P, Q \in \mathbb{P}$. It is strictly proper if the equality holds if and only if $P=Q$.

\hfill
\subsection{Construction of Proper Scoring Functions}
\hfill\\

We know, from \cite{guo2003}, that a characterization of a Bregman Divergence is that:

\begin{equation}\label{bregeq}
    \underset{Q \in G}{\text{argmax  }} \mathbb{E}[-D_f(P, Q)] = \mathbb{E}[P|G]
\end{equation}

But, for a scoring function, where the forecaster reports Q, and his true probability measure is P, we have that:

\begin{equation*}
    \underset{Q \in \mathbb{P}}{\text{argmax  }} \mathbb{E}[S(\mathbbm{1}_{\omega}, Q)] = \mathbb{E}[\mathbbm{1}_{\omega}|\mathcal{F}] = P
\end{equation*}

Hence, it is always optimal for a forecaster to report his true density function P. Thus, a proper scoring function is equal to the negative of some Bregman Divergence. But we know \cite{b2005} that to construct a Bregman Divergence, we only need to pick a strictly convex function f and write:

\begin{equation*}
    S(P, Q) = f(p) - f(q) - \langle \nabla f(q), p-q \rangle
\end{equation*}

where $\nabla f(q)$ denotes the sub-gradient of f at q.

\hfill
\section{Models}
\hfill

\subsection{CAPM Model}
\hfill\\

Our model follows the intuition of the Nobel prize winning Capital Asset Pricing Model introduced by Treynor \cite{treynor1961market} \cite{treynor1962jack} and Sharpe \cite{sharpe1964capital}. We treat each state as a stock with the national popular vote playing the role of the market. To calibrate this model, we consider over 1000 polls that are obtained from RealClearPolitics and model each state's popular vote as a GAM with the single factor being the national polls. Once each state has been calibrated, we treat the market as a continuous time stochastic process which can take any number of values over the remaining period. Simulating from this model and for each state specific noise, we obtain a number of possible scenarios for each state. The computation of the electoral votes won by each candidate generates a winner for each scenario.

Consider an election with two candidate $c_i: i \in \{1, 2\}$. At each time $t$ we denote the popular vote spread of candidate $c_1$ in each state by $S^i_t \text{ for } i \in [1, 50]$ and denote by $M_t$ the national popular vote spread (at $t$). For each state, we assume that the spread follows the model:

\begin{equation} \label{eq:capm}
S^i_t = \beta^i \phi(M^i_t) + \epsilon^i
\end{equation}

where $\phi$ is some well-behaved function (in our case a non-parametric regression).

Note, that unlike the traditional CAPM model, our equation is in levels rather than returns. This is mainly done to avoid the noise created by multiple polls which are very close to each other. We also make the assumption that the $\beta^i$ in \ref{eq:capm} does not depend on time, although a Kalman filter type methodology could easily be used to incorporate this into the model. We model the national popular vote in the following way:

\begin{equation} \label{mkt}
dM_t = (\sigma^{\text{samp}} + \sigma^M) dW_t 
\end{equation}

where $W_t$ denotes a canonical Brownian motion defined on the space $\Omega \equiv \mathcal{C}_0 ([0, T], \mathbb{R})$, where $T$ denotes the time of the election. We further assume that the white noise $\epsilon^i \sim \mathcal{N}(0, \sigma^i)$.

We use Ordinary Least Squares Regression to calibrate the $\beta^i$ for each state and analyze the standard deviation of the residuals to obtain $\sigma^i$. The data methodology is reviewed in Section \ref{data-section}. We use the standard deviation of the national spread as our best estimate for $\sigma^m$. To forecast, we simulate 10000 paths of the brownian motion $W_t$ and for each path we simulate 50 state specific noises $\epsilon^i$. Doing this allows us to obtain a popular vote estimate for each state $i$ in each simulation $j \in [1, M]$ which we denote by $S^{ij}_T$. Converting the popular vote in each state to an electoral vote enables us to obtain a winner for each election.

We define the probability of winning each state (i) as:

\begin{flalign*}
   && \mathbb{P}^{c_1}_i := \frac{\sum\limits^M_{j=1} \mathbbm{1}_{S^{ij}_T > 0.5}}{M} && \text{(for $i \in [1, 50]$)}
\end{flalign*}

Denote by $\text{Elec}^j$ as the total number of electoral votes for candidate 1 in simulation $j$. Thus $\text{Elec}^j: S^j_T \rightarrow \mathbb{N}$ and denote $\mathbb{P}^{c_1}$ as the probability of candidate $c_1$ winning the election, where:

\begin{equation}
    \mathbb{P}^{c_1} := \frac{\sum\limits^M_{j=1} \mathbbm{1}_{\text{Elec}^j \geq 270}}{M}
\end{equation}

The time series of probabilities can be seen in Figure \ref{capm-ts}. The simulations can be seen in Figure \ref{capm-sim}.

\newpage

\subsection{Bayesian Methods}

\subsubsection{Robust Regression}

For our first advanced model, we assume a similar CAPM structure. However, we additionally postulate that:

\begin{align*}
    \alpha^i &\sim \N(\alpha^{OLS}, \sigma^{\alpha}) \\
    \beta^i &\sim \N(\beta^{OLS}, \sigma^{\beta}) \\
    \sigma^i &\sim |\N(0, \sigma^{\epsilon^i})|\\
    \mu^i &= \alpha^i + \beta^i M_T \\
    S^i_T &\sim \text{StudentT }(\mu^i, \sigma^i, \nu=3)
\end{align*}

Where $\sigma^{\alpha} = 0.01$ and $\sigma^{\beta} = 1.$. Finally, we simulate $M_T$ as a Brownian Motion, and draw from the predictive distribution of $S^i_T$ for the state-specific noise. An example of this can be seen in Figure \ref{nevada-ts}. The simulations can be seen in Figure \ref{student-sim}.

\subsubsection{Hierarchical Regression}

\begin{figure}[h]
\begin{center}
\includegraphics[width=14.5cm, height=7cm]{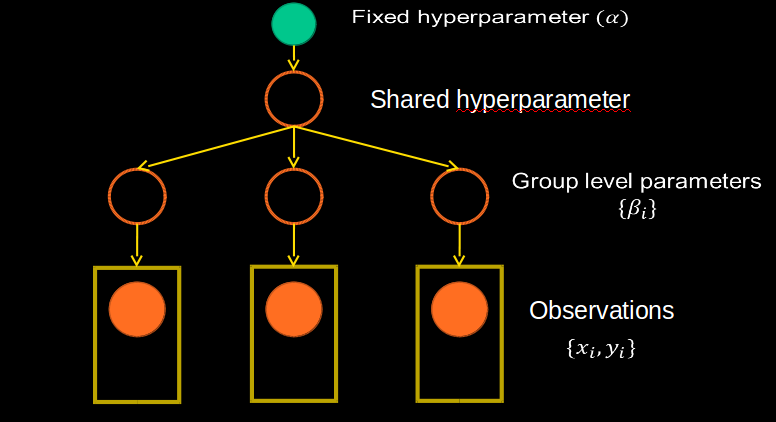}
\caption{Using the Bayes Ball Algorithm we can see that each node is influenced by the others through the posterior distribution of the shared hyperparameter.}\label{graph-struc}
\end{center}
\end{figure}

A potential model, is to assume the Graphical Structure in Figure \ref{graph-struc}.

We can sample our State Level noise from this model, and use the Bachelier Process to simulate the Market.

\section{Evaluation of Forecasters}

The traditional way of evaluating a forecaster is the Brier Score. Which is defined, for a realizations $\omega_t$ and a sequence of probabilities $p_t$ as:

\begin{equation}
    \text{Brier(T)} = \sum^T_{t=1} (\omega_t - p_t)^2
\end{equation}

The Brier Score is proper \cite{savage1971} and the lower the value the better. A perfect forecaster would have a Brier Score of 0.

However, we consider a simple example, where two forecasters are asked to give a probability an event $o$, whose true probability is $\frac{1}{10}$, will happen every day. Forecaster A provides a constant probability of $0$ while forecaster B provides a constant probability of $\frac{1}{4}$. If both provide the same constant probability for 100 time periods, Brier(A) = 1, while Brier(B) = 2.25. It is not clear whether providing a zero probability for a low probability event is optimal. Thus, we need better tail event behaviour. This motivates the log-likelihood.

\begin{equation}
    \text{Log Likelihood(T)} = \sum^T_{t=1} log(\omega_t p_t + (1-p_t)(1-\omega_t))
\end{equation}

As in \ref{shape-fig}, we see that the Log Likelihood goes parabolically to infinity at the tails, while the Brier/Selten Scores become flat.

\hfill
\subsection{Density Comparison}
\hfill\\

Both the Brier and the Selten score however provide an incomplete picture. Consider the US Election of 2016, where the two most publicized forecasters were Nate Silver and Dr. Sam Wang. Nate Silver gave Hillary Clinton a (~70\%) chance to win, while Dr. Wang gave her $> 99.999\%$. While the realization was 232 EV for Clinton, a question begs a subtle response, what if Clinton had won in a Reagan-esque landslide (say with $\geq 489 EV$), the Brier and Log Likelihood would say that Dr. Wang was the better forecaster. However, a look at the histogram begs to differ:

\begin{figure}[!tbp]
  \centering
  \subfloat[Princeton Election Consortium]{\includegraphics[width=\textwidth]{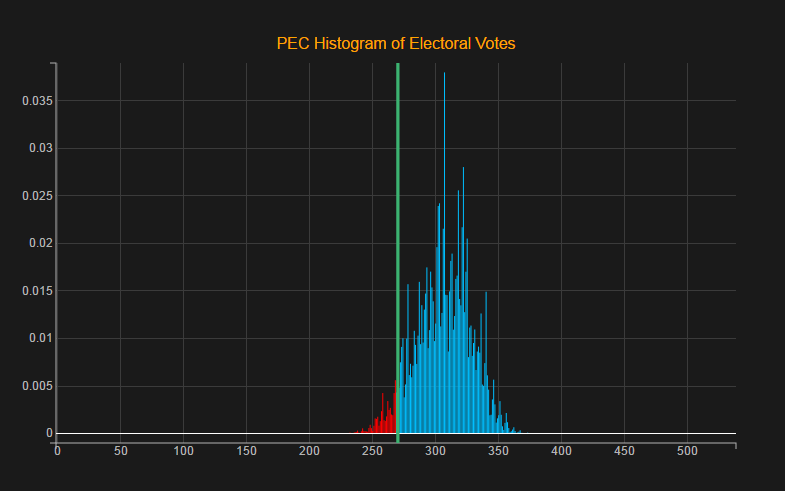}\label{fig:f1}}
  \hfill
  \subfloat[FiveThirtyEight]{\includegraphics[width=\textwidth]{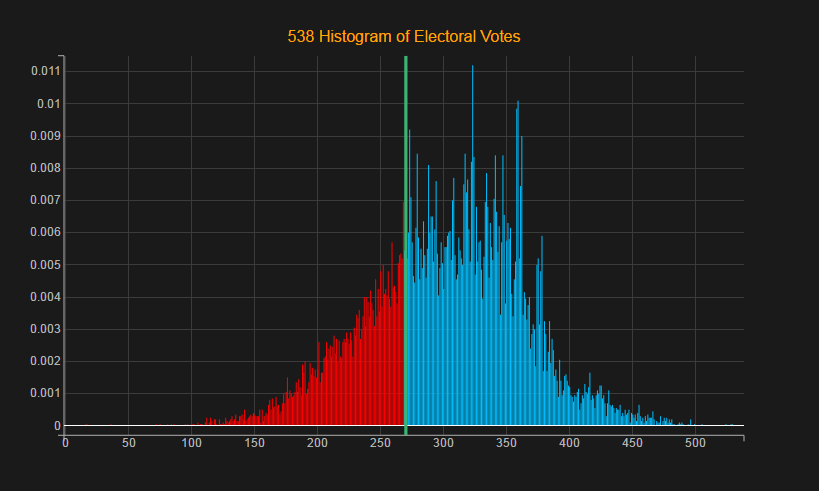}\label{fig:f2}}
  \caption{The Histograms reveal that 538 had assigned more probability to a Clinton Landslide than PEC.}
\end{figure}

We propose two methods to evaluate forecasters based on their histogram.

\hfill
\subsubsection{Selten Score}
\hfill\\

The first, the Selten Score \footnote{Named after Nobel Laureate Reinhard Selten. Though it was inspired by his argument for the singularity of the Brier Score, we could not find a reference which uses the Brier Score in this way.} \cite{selten}, is equivalent of taking the Brier Score in each bin of the Histogram. We treat each bin independently. We know from \cite{rudin} that the maximum of a sum of functions is equivalent to the sum of maximums. As a result, the Selten Score is proper.

For $N$ bins in the Histogram, each assigned a probability $p_i$, with a realization $\omega \in {i^*}$, we have:

\begin{align*}
    \text{Selten Score} &= 1 - \sum^N_{i=1} (\omega_i - p_i)^2 \\
    &= 2p_{i^*} - \sum\limits^N_{i=1} p^2_i
\end{align*}

We interpret this as a score that rewards assigning a high probability to the correct bin ($2 p^*_i)$ and penalizes the forecaster for having too spread out of a distribution ($\sum^N_{i=1} p^2_i$).

\hfill
\subsubsection{CDF Score}
\hfill\\

The Selten Score takes no account of the topology of the different bins. In order to account for this topology, we propose to use a different scoring function, which looks at the Brier Score above or below each level. The Daily Kos methodology that uses Binomial Models as priors, lends itself to producing spikes in the EV Histogram (see \ref{dk-histogram}). Thus even though the results are similar to the Princeton Election Consortium (see Figure \ref{pec-hist}), the Selten \ref{overall-selten} Score penalizes it as the topology of the bins is not factored into the calculation. Two forecasters who gave the the entire probability mass to a single bin (say 226 and 538) would have the same Selten score, even though the result was 227 EV. We seek a scoring function that factors this topology into account.

\begin{equation*}
    \text{CDF Score}(F, \omega) = \int (F(x) - \mathbbm{1}_{x \geq \omega})^2 dx
\end{equation*}

A proof for the propriety of this scoring function is given in Appendix A.
\\
\textbf{The results for each model on each scoring function is given in Appendix C.}

\subsubsection{Comparison of Different Scoring Functions}

As seen in Figure \ref{score-comp}, the behaviour of diffferent scoring functions becomes very important in the tails. While the Selten/Spherical/Brier become flat, the log moves parabolically to $-\infty$, which makes it hypersensitive to low probability events, while the Selten/Spherical/Brier are undersensitive. Once again, we see the benefit of the CDF Score, as it moves linearly to $-\infty$.

\hfill
\section{Trading Score and Online Learning}
\hfill

All of the forecasts given above are ex-post forecasts, i.e., they require the realization of an event. However, in many cases, such as Election Modelling, we would like to judge the efficacy of forecasters before the realization of the event so that we may combine them to create an optimal mixture of forecasts.

For this purpose, we propose the trading score. Consider 2 forecasters predicting a binary event, we assume each posts a time series of forecasts $a_t$ and $b_t$ respectively. Assume each day each forecasters take a position that is proportional to their distance from a reference level $s_t$. This can be the betting market (see Appendix C for results) or the average of the 2 forecasters. Given that each forecaster takes a position everyday (bought at the betting market or at the average), we can re-evaluate the value of this position everyday and treat the cumulative P\&L as an online scoring function. Finally, on a realization of an event, we either settle at the betting market (which converges to the 0 or 1) or at the realization of (0 or 1). See Appendix B for a proof.

Using the trading score, with the betting market as a reference, we employ the weighted majority algorithm, with the trading score and the quadratic loss (brier score) as the Loss function in each case. At each date, we take a position relative to our prediction and then backtest this position over all available time periods. Finally, we look at the entire profit and loss of the algorithm that buys and sells the betting market as its predicted price.

For $N$ experts, our weights are initialized $w_{0, i} = \frac{1}{N}$, for $i \in [1, N]$, and updated as follows:

\begin{equation*}
    w_{t+1} \leftarrow w_t e^{-\eta L(\hat{y}_{t,i}, y_{t})}
\end{equation*}

with the prediction at each round being:

\begin{equation*}
    \hat{y}_t = \frac{\sum\limits^N_{i=1} w_{t,i}y_{t,i}}{\sum\limits^N_{i=1} w_{t,i}}
\end{equation*}

where L denotes the cumulative loss of each expert. \cite{mohri12} derive a bound on the Regret upto time T and show that for

\begin{align*}
    \eta &= \sqrt{\frac{8 \log N}{T}}\\
    \implies \text{Regret }(T) &\leq \sqrt{\frac{T}{2}\log N}
\end{align*}

\newpage

\section{Data Methodology for the 2016 Presidential Election} \label{data-section}

The data for both the national and state polls is obtained from RealClearPolitics. The for each polls consists of the following attributes:

\begin{itemize}
    \item Name of the Pollster
    \item State (= US for National Polls)
    \item Sample Size
    \item Sample Type $\in$ [Registered Voters, Likely Voters, All]
    \item Trump \%
    \item Clinton \%
    \item Johnson \%
    \item Stein \%
\end{itemize}

Using this data, we assume WLOG that $c_1$ = Clinton \%  and proceed to use $S^i_t = \text{Clinton}^i_t - \text{Trump}^i_t \%$ as the primary variable to model. We choose to model only the Democratic and Republican candidate as the leading independent Gary Johnson shows his highest poll \% in a New Mexico poll conducted by the Albuquerque Journal \cite{albjournal}, where Clinton still leads by 11 \%. Though it should be noted here that Johnson winning New Mexico can create many more scenarios of an electoral deadlock, in which case the state popular vote is no longer an accurate way to forecast the winner of the election. A more detailed methodology would allow for a fat tail where this is possible.

\subsection{Missing States}

Numerous states are either not featured in the poll database or contain fewer than 4 points, making any inference with them extremely dubious.

The following states lacked sufficient data to do any analysis:

\begin{multicols}{3}
\raggedcolumns
\begin{itemize}
    \item Alabama
    \item Alaska
    \item Hawaii
    \item Kentucky
    \item Montana
    \item Nebraska
    \item North Dakota
    \item Oklahoma
    \item South Dakota
    \item Tennessee
    \item West Virginia
    \item Wyoming
    \item Washington D.C.
\end{itemize}
\end{multicols}

As a result, to compound the data we used the state's popular vote spread for the party associated with candidate $c_1$ in each election going back to 1976 along with the national popular vote spread for that year.

\subsection{Asynchronity in State and National Polls}

As a result of sparse data in each state, there may not be state and national polls available in each time period, or alternately there may be multiple national polls within the same period. To deal with this we calibrate a non-parametric regression with a Gaussian Kernel to the national poll. The chosen bandwidth is 5.0, which is determined empirically. 

This methodology may be criticized for its naivety. And in fact, most other public methodologies \cite{silver2016} weight each poll by some measure of it's quality. However, a recent paper \cite{pasek2015predicting} surveys the long history of aggregating polls and questions whether weighting by accuracy measures is relevant when the error arises from multiple sources (not just sampling error). The paper posits that it may be just as, if not more, appropriate to aggregate using a simple weighted average.

\begin{figure}[h]
\centering
\includegraphics[width=10cm, height=7cm]{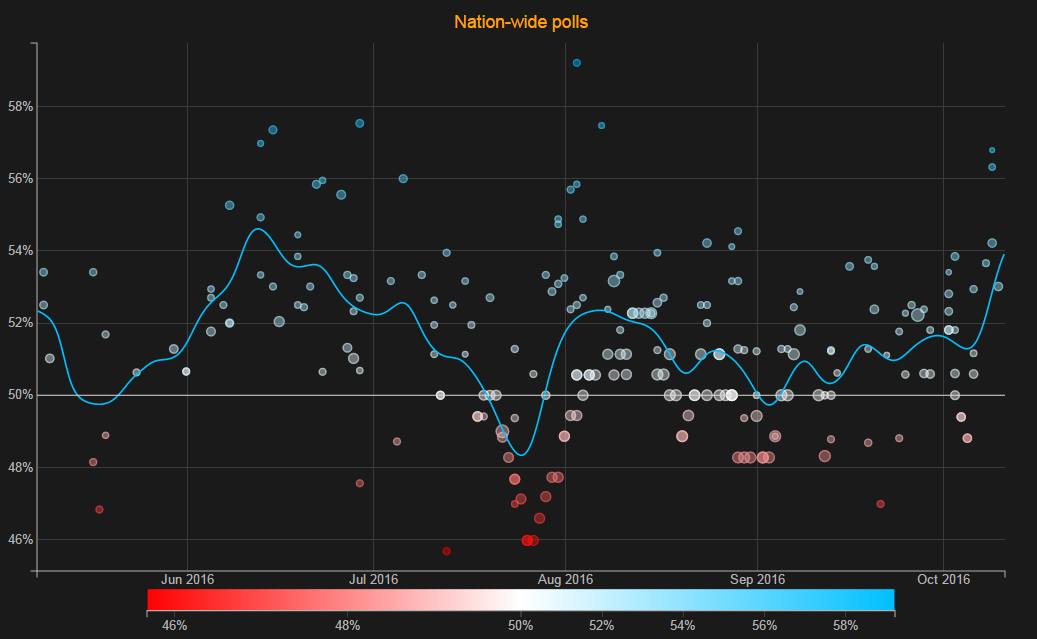}
\caption{Plot of the non-parametric regression for the Clinton (Candidate C1) Spread. The dots represent the actual polls, while the line represents the smoothed aggregation used.}
\end{figure}

\subsection{Modeller Data}

In order to compile the experts, we scraped the forecasts made publicly available for the following forecasters:

\begin{multicols}{2}
\raggedcolumns
\begin{itemize}
    \item FiveThirtyEight (Now/Plus/Polls' Models)
    \item Huffington Post
    \item New York Times
    \item Princeton Election Consortium
    \item Daily Kos
\end{itemize}
\end{multicols}

Additionally, we use the CAPM and Option Market Models we have created as our 5th and 6th experts. We use the data provided by \cite{lott}.

\section{Results}

As we see in Figure \ref{overall-brier}, the CAPM and the Online Learning Model performed better than everyone except the FiveThirtyEight, the Betting Market, the CAPM and the Option Market Model. The additional noise in the CAPM Model aslo helped it perform very well. In a state average, we see that predicting OH, FL, IA, NC and AZ correctly from early on made the CAPM perform very well. An Electoral Vote weighted average shows that the CAPM performed best.\footnote{It is not clear that advanced Bayesian methods are helpful here. The probabilities provided by the Student T regression as well as the Hierarchical Model were very similar to the CAPM Model. If we study the histogram in Figure \ref{student-sim}, it is not clear that fat-tailed distributions are relevant here, unless we believe there are non-trivial probabilities of a landslide.}

FInally, the log damages the Princeton Election Consortium as they reported 100\% probabilities for Clinton winning. The same applies for Daily Kos in the State-wise and EV weighted averages. 

We see the flaw in the Selten Score \ref{overall-selten} as the DK Histogram in Figure \ref{dk-histogram} has spikes as an artifact of the modelling technique, which the Selten ignores. The CDF Score \ref{overall-cdf} on the other hand shows that the two are really comparable.

The Online Learning Algorithm performs very well, \ref{pl-ts} but is unable to overcome the fact that most experts were sure about Clinton and there was a strong discontinuity in the betting market towards the end. In terms of Scoring Functions, it performed much better than most algorithms, and was comparable to the CAPM Model. 

The trading score performed much better than a conventional quadratic score (see Figure \ref{pl-nolast}) but was more bullish than the quadratic score (see Figure \ref{pl-ts-quad}). As a result, the final P\&L was slightly worse. The MSE (see Figure \ref{mse}) for predicting the betting market was lower for the Online Algorithms than for any modeller except 538 \footnote{As a footnote, its not clear that the betting markets, like all of us are not just following 538.} We once again see that there are some issues to be sorted out, while the trading score does solve the ex-ante, ex-post problem, it does not provide a good reference when no market is available. Martingality of prices by the No-Arbitrage Theorems imply that the average is not a good reference\footnote{Since by linearity of expectation, taking an equally weighted average should be ideal, and by the Martingale property the best guess of the future value is the current value.} 

The option methodology does not have a clean way to select securities. Trump's rhetoric allowed us to interpret the USDMXN peso correctly, but there exists no clean way to select securities. This would be an ideal task for further machine learning research.

Finally, to improve the Online Learning Algorithms, better scoring functions are needed. The CDF score seems like a great candidate, though data for experts is very hard to get. We hope that this paper will motivate the use of density scoring versus simple Brier or Log-Likelihood Scoring. Currently very few forecasters report the density and fewer keep a catalogue of it. Providing this data will allow for a better aggregation of the different forecasters, and better model building.

\bibliographystyle{abbrv} 
\bibliography{fml2016-sample}

\appendix

\newpage

\section{CDF Score is Proper}

Consider a forecaster, whose true measure is $\mathbb{P}$ and who reports $\mathbb{Q}$.Then his expected score is:

\begin{align*}
    S(\mathbb{P}, \mathbb{Q})&= \expp \biggl[ \int\limits^{\infty}_{-\infty} (Q(y) - \mathbbm{1}_{y \geq x})^2 dy \biggr] \\
    &= \expp \biggl[ \int\limits^{\infty}_{-\infty} Q^2(y) dy \biggr] - 2 \int\limits^{\infty}_{-\infty} Q(y) \expp [ \mathbbm{1}_{y \geq x} ] dy + \int\limits^{\infty}_{-\infty} \expp [\mathbbm{1}_{y \geq x}] dy \\
    &= \int\limits^{\infty}_{-\infty} Q^2(y) dy - 2 \int\limits^{\infty}_{-\infty} Q(y) P(y) dy + \int\limits^{\infty}_{-\infty} P(y) dy \\
\end{align*}

Differentiating with respect to $\mathbb{Q}$ and setting the derivative $ = 0$, and assuming regularity conditions on the densities, we get:

\begin{align*}
    \frac{\partial S(P, Q)}{\partial Q} &= 2 \int\limits^{\infty}_{-\infty} Q(y) dy - 2 \int\limits^{\infty}_{-\infty} P (y) dy \\
    \implies \int\limits^{\infty}_{-\infty} P(y) dy &= \int\limits^{\infty}_{-\infty} Q (y) dy \\
    \implies P(y) &= Q(y) \text{ almost everywhere }
\end{align*}

Now, we consider a Cramer type of divergence function:

\begin{equation*}
    d(P, Q) = \int\limits^{\infty}_{-\infty} (P(y)-Q(y))^2 dy
\end{equation*}


\begin{equation}
    \underset{Q}{\text{argmax }} \expp \biggl[ \int\limits^{\infty}_{-\infty} ( \mathbb{P}(y) - Q(y) )^2 dy \biggr] = \expp \bigl[ \mathbb{P} ]
\end{equation}

which shows that the CDF score is proper.

\newpage

\section{The Trading Score is Proper}

Consider, a two date one period model. The position at time t is $(a_t - b_t)$, which is bought at a price $\frac{a_t + b_t}{2}$. At time T, we settle at the realization, which gives us

\begin{align*}
    S(a_0, T) = \mathbb{E}^{\mathbb{P}} [\text{PNL}(a_0, T)] &= \mathbb{E}^{\mathbb{P}} [-\mathbbm{1}_{\omega} (a_0 - b_0) + \frac{a^2_0 + b^2_0}{2}] \\
    \implies \frac{\partial S}{\partial a_0} &= \mathbb{E}^{\mathbb{P}}[-\mathbbm{1}_{\omega} + a_0] \\
    \implies \underset{a_0}{\text{argmax  }}\text{PNL}(a_0, T) &= \mathbbm{E}^{\mathbb{P}}[\mathbbm{1}_{\omega}]
\end{align*}

Hence, the profit and loss of this trading score corresponds to a strictly proper scoring function. A simple induction argument proves this for N time periods. For the position take at period $T-1$, we consider maximizing the profit and loss.

\begin{align*}
    S(a_{T-1}, T) = \mathbb{E}^{\mathbb{P}} [\text{PNL}(a_{T-1}, T)] &= \mathbb{E}^{\mathbb{P}} \biggl[-\mathbbm{1}_{\omega} (a_{T-1} - b_{T-1}) + \frac{a^2_{T-1} + b^2_{T-1}}{2} | \mathcal{F}_T\biggr] + \\ &PNL(a_{0 < t < T-1}, T)\mathbb{E}^{\mathbb{P}}[\mathbbm{1}_{\omega}|\mathcal{F}_T]\\
    \implies \frac{\partial S}{\partial a_{T-1}} &= \mathbb{E}^{\mathbb{P}}[-\mathbbm{1}_{\omega} + a_{T-1} | \mathcal{F}_T] \\
    \implies \underset{a_{T-1}}{\text{argmax  }}\text{PNL}(a_{T-1}, T) &= \mathbbm{E}^{\mathbb{P}}[\mathbbm{1}_{\omega} | \mathcal{F}_T]
\end{align*}

Now, we can do the same for $a_{T-2}$, setting $\mathbb{E}[a_{T-1} | \mathcal{F}_{T-2}] = \mathbb{E}^{\mathbb{P}}[\mathbbm{E}^{\mathbb{P}}[\mathbbm{1}_{\omega} | \mathcal{F}_T] | \mathcal{F}_{T-2}] = \mathbb{E}^{\mathbb{P}} [\mathbbm{1}_{\omega} | \mathcal{F}_{T-2}]$, by the tower property of the conditional expectation \cite{mohri12}.

\newpage

\section{Results of Models}\label{res-sec}

\begin{figure}[h]
\begin{center}
\includegraphics[width=16.5cm, height=7cm]{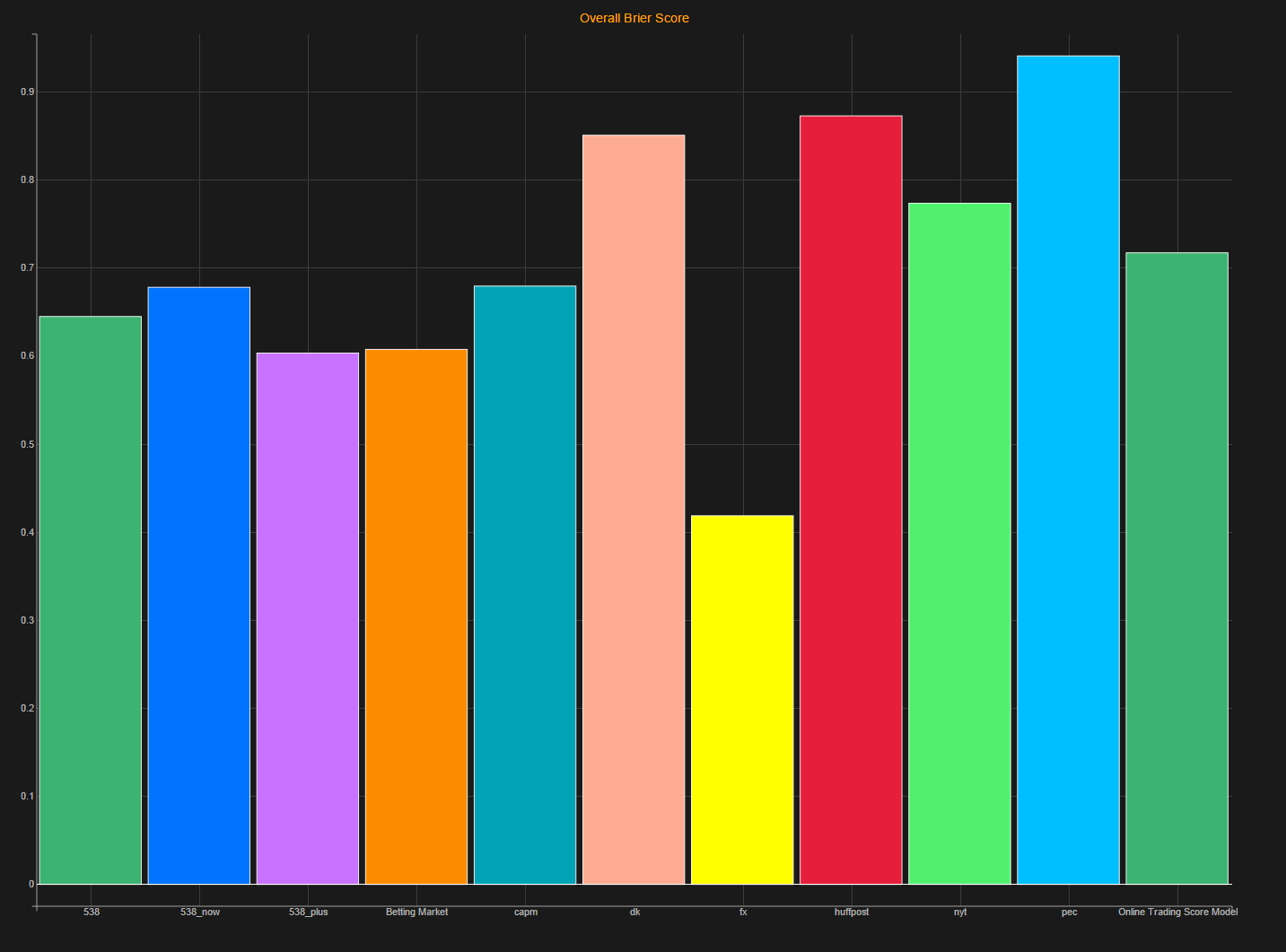}\label{overall-brier}
\caption{In the overall Brier we see that the Online Model performed better than all the bullish models.}
\label{overall-brier}
\end{center}
\end{figure}

\begin{figure}[h]
\begin{center}
\includegraphics[width=16.5cm, height=7cm]{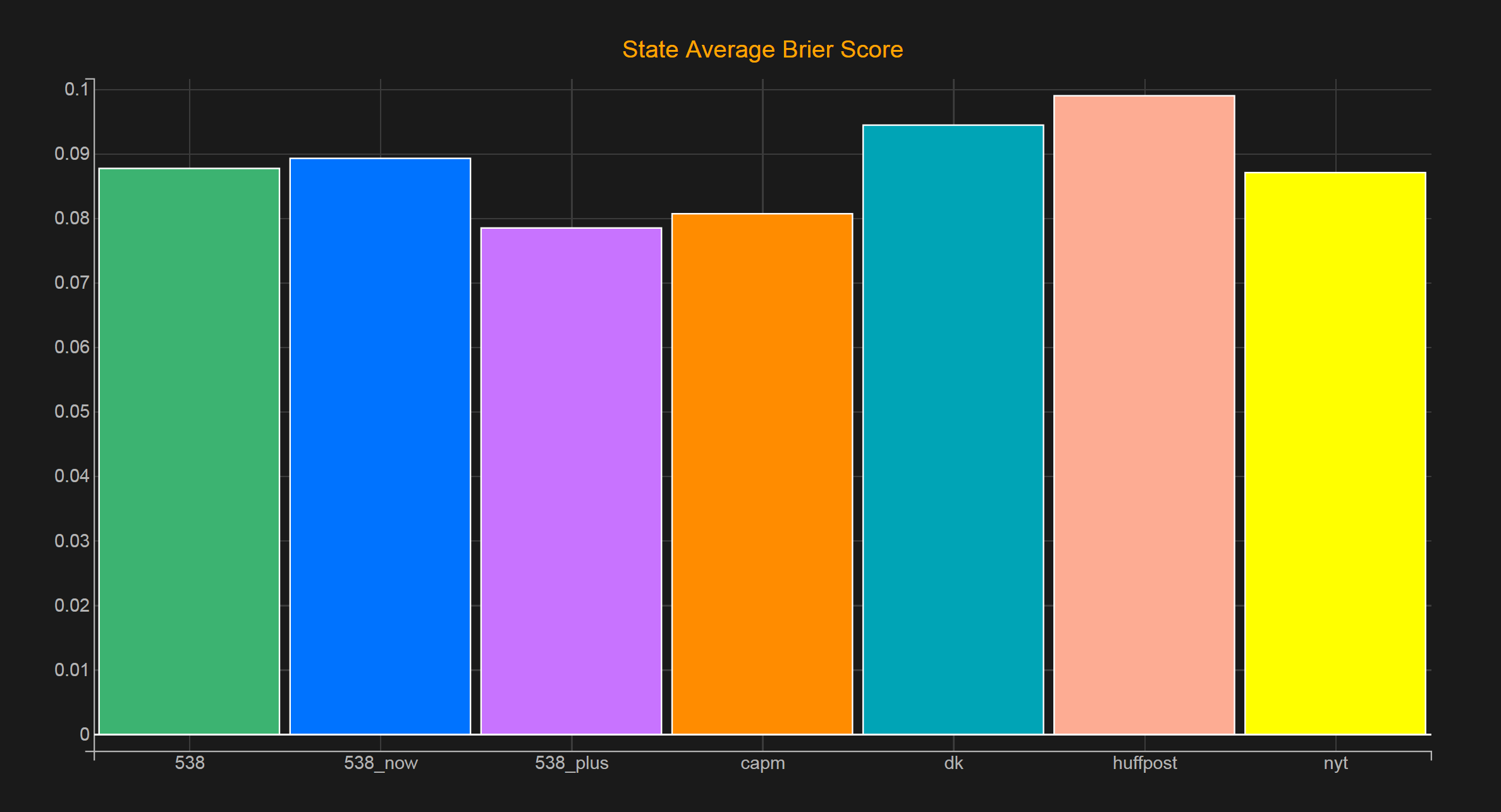}
\caption{The CAPM Model did better than all models except the FiveThirtyEight Polls Plus Model.}
\label{state-brier}
\end{center}
\end{figure}

\begin{figure}[h]
\begin{center}
\includegraphics[width=16.5cm, height=7cm]{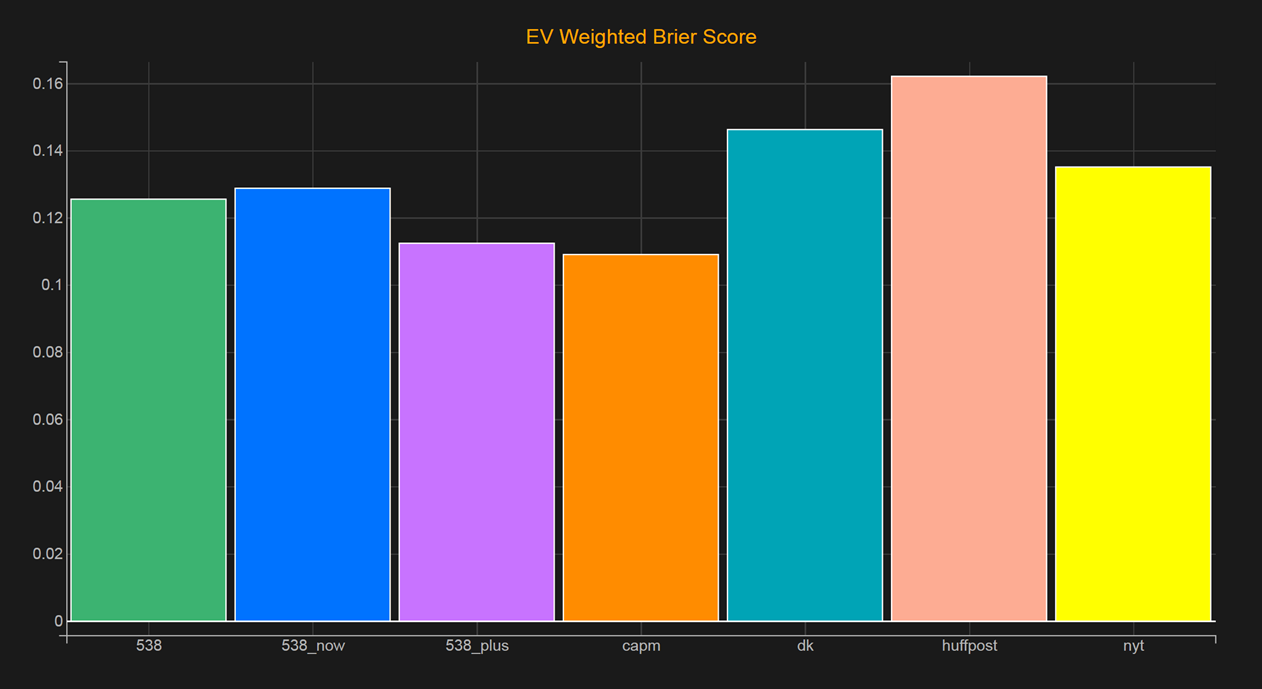}
\caption{The CAPM Model correctly predicted OH, FL, IA, NC from very early on, which allowed better scores than FiveThirtyEight when weighted by EV, making it the best EV weighted model.}
\label{ev-brier}
\end{center}
\end{figure}

\begin{figure}[h]
\begin{center}
\includegraphics[width=16.5cm, height=7cm]{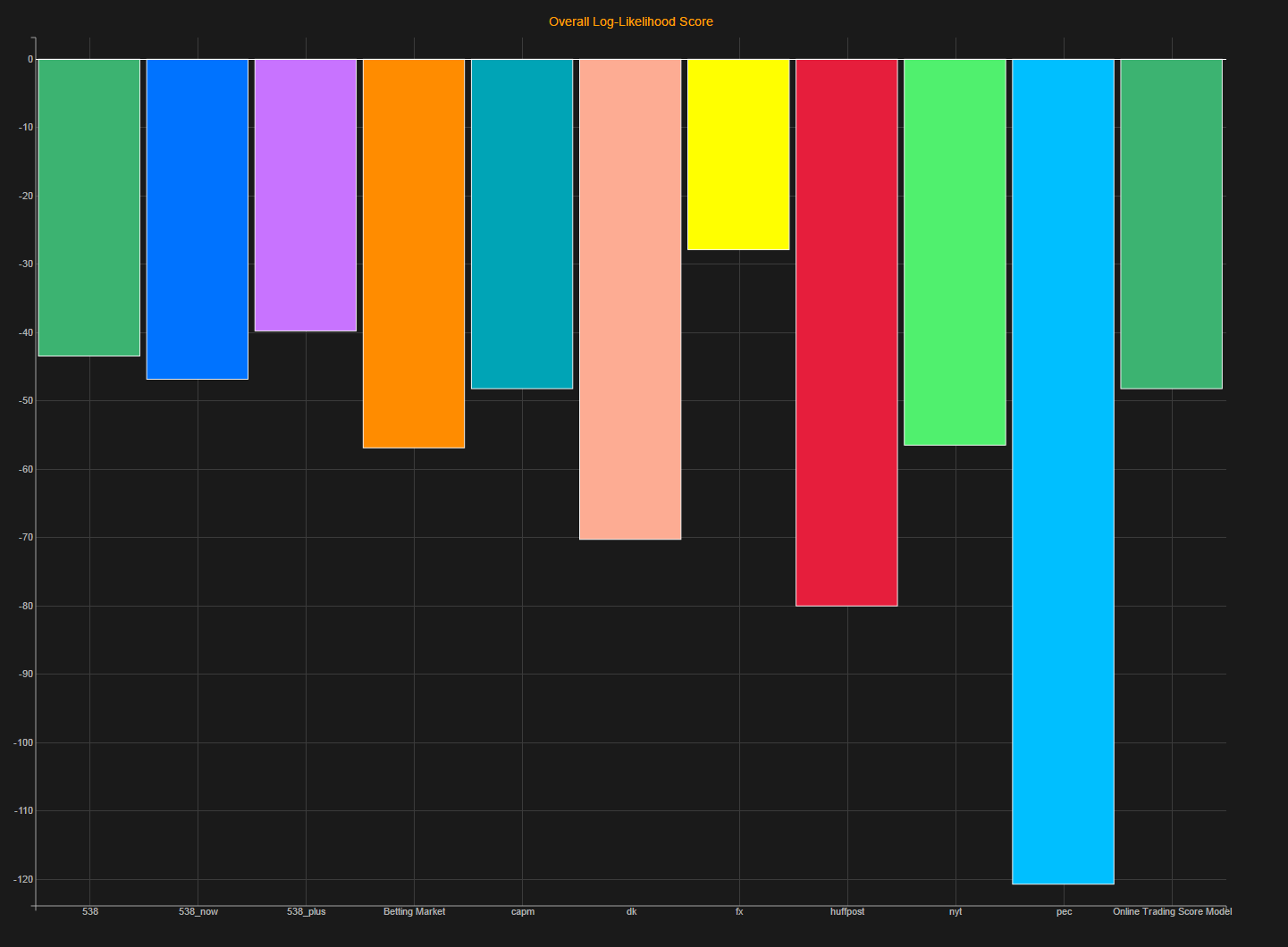}
\caption{Princeton Election Consortium and HuffPost were heavily penalized for reporting probabilities close to 100\%, due to the hypersensitivity of the log \cite{selten}.}
\label{overall-log}
\end{center}
\end{figure}

\begin{figure}[h]
\begin{center}
\includegraphics[width=16.5cm, height=7.75cm]{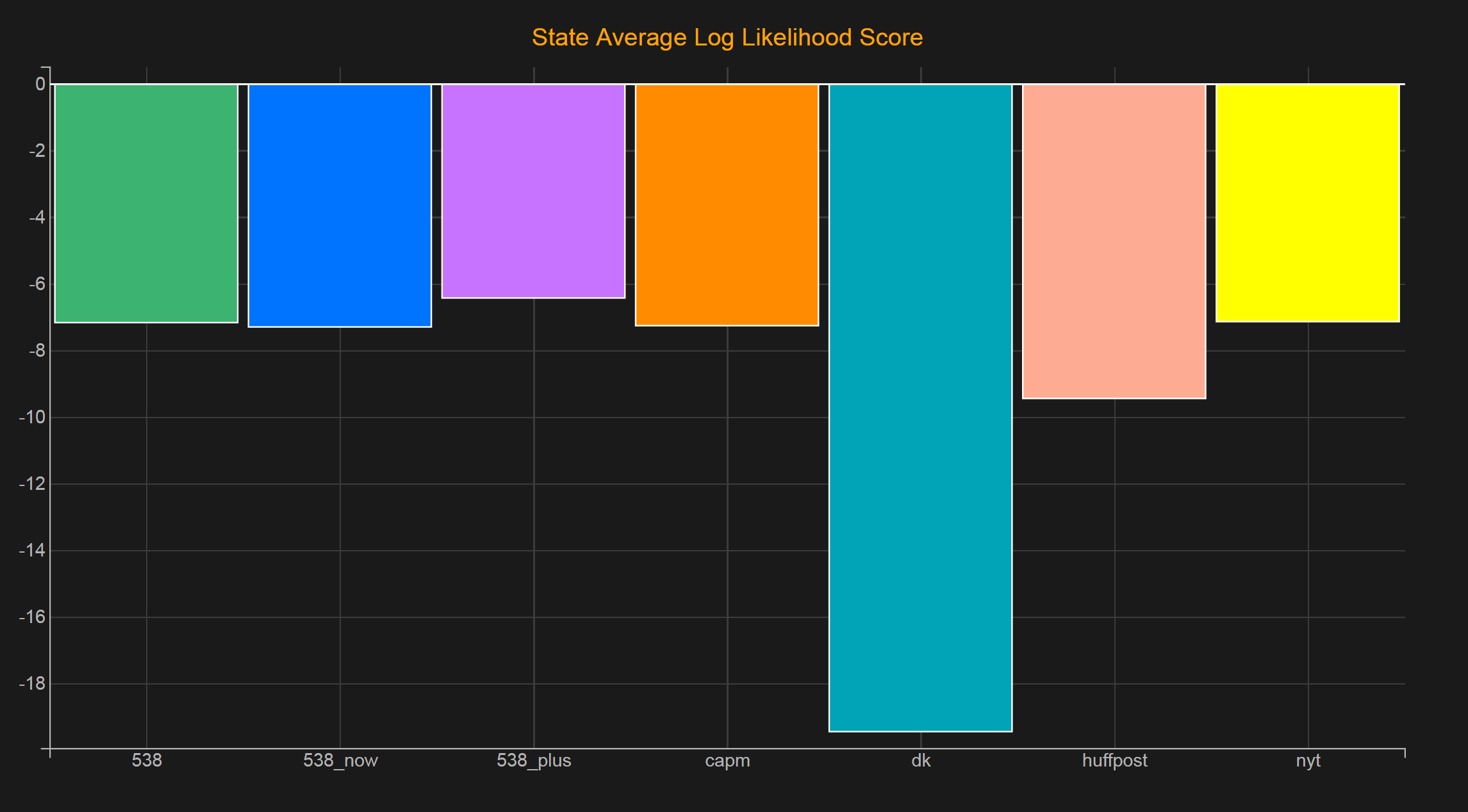}
\caption{Daily Kos was very sure about MI, PA, WI, causing a massive loss.}
\label{state-log}
\end{center}
\end{figure}

\begin{figure}[h]
\begin{center}
\includegraphics[width=16.5cm, height=7.75cm]{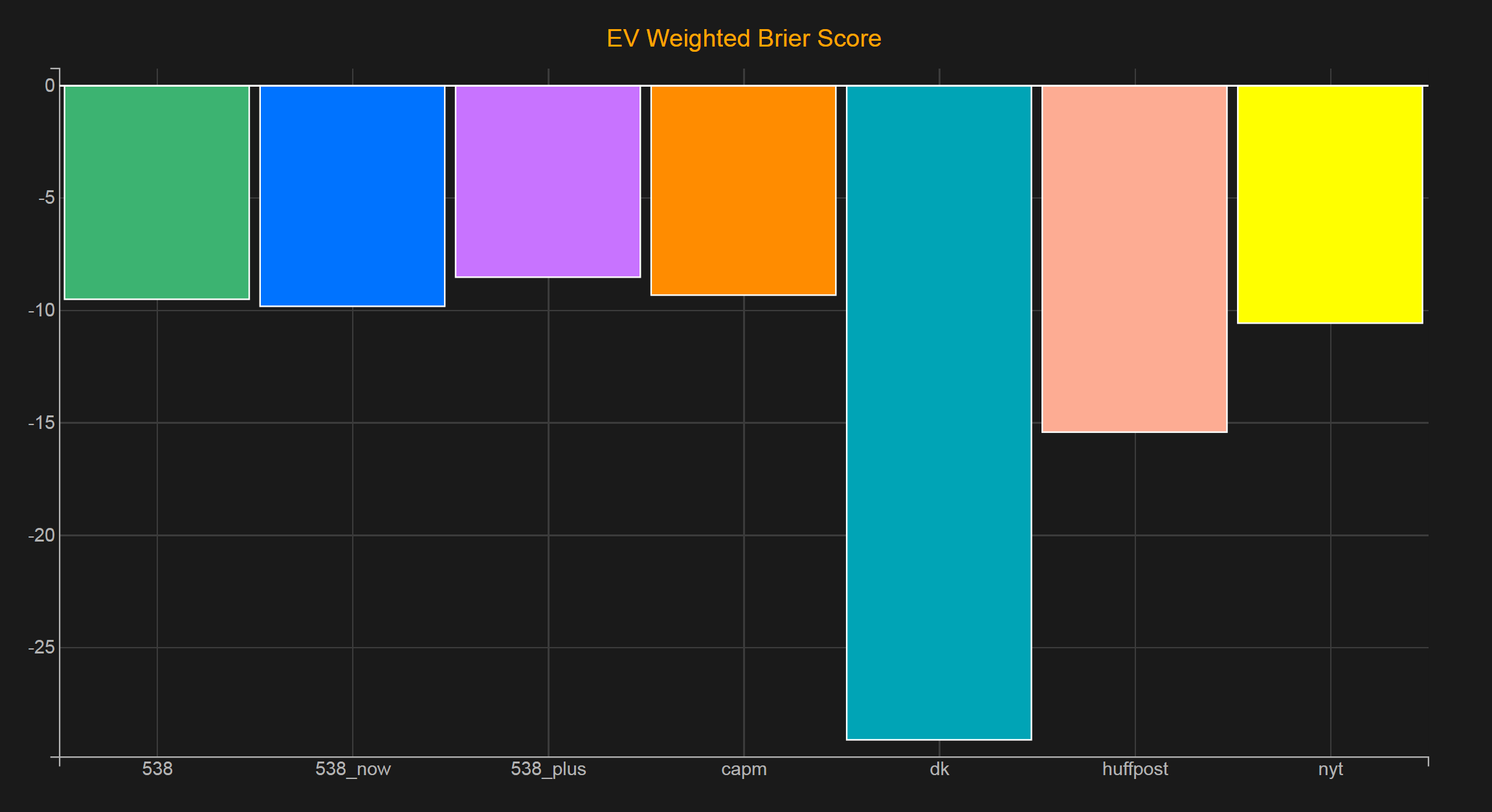}
\caption{EV Weighted Log Likelihood Score: Daily Kos was very sure about MI, PA, WI, causing a massive loss.}
\label{ev-log}
\end{center}
\end{figure}

\begin{figure}[h]
\begin{center}
\includegraphics[width=16.5cm, height=7.2cm]{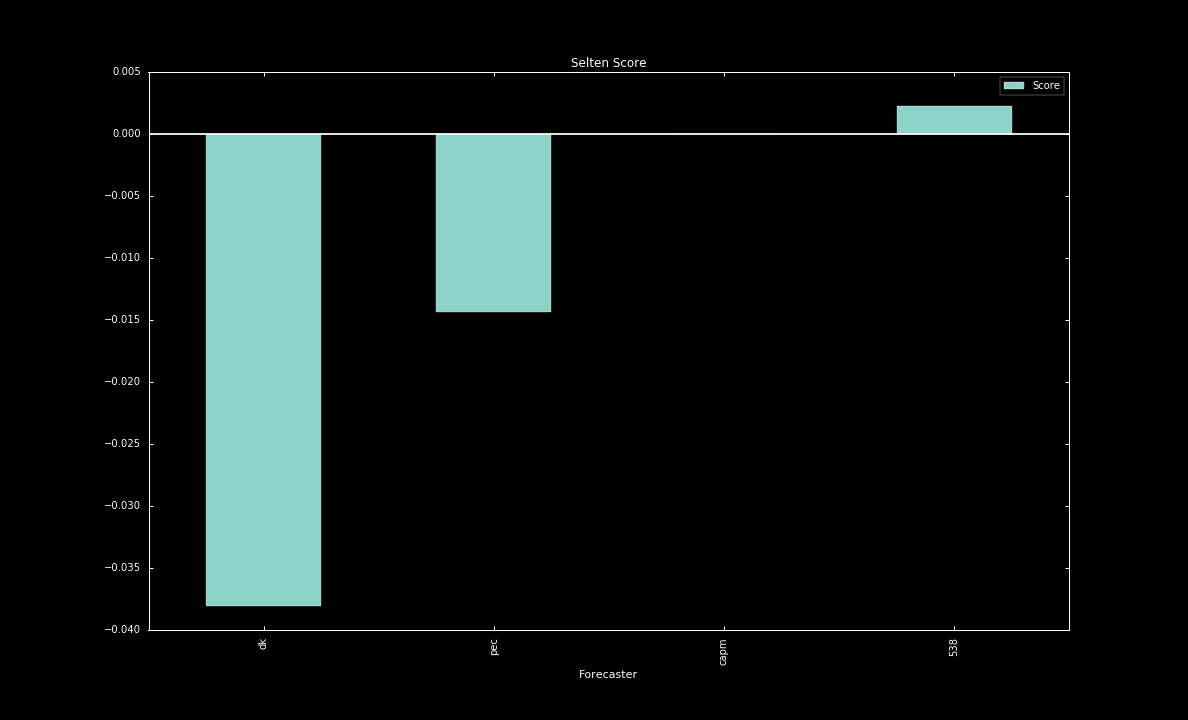}
\caption{The DK and the PEC histograms were very narrow. Here we see the flaw of the Selten, as DK's logit model created spikes in the Histogram close to 232 that are not accounted for, giving it the lowest Selten Score}
\label{overall-selten}
\end{center}
\end{figure}

\begin{figure}[h]
\begin{center}
\includegraphics[width=16.5cm, height=7.2cm]{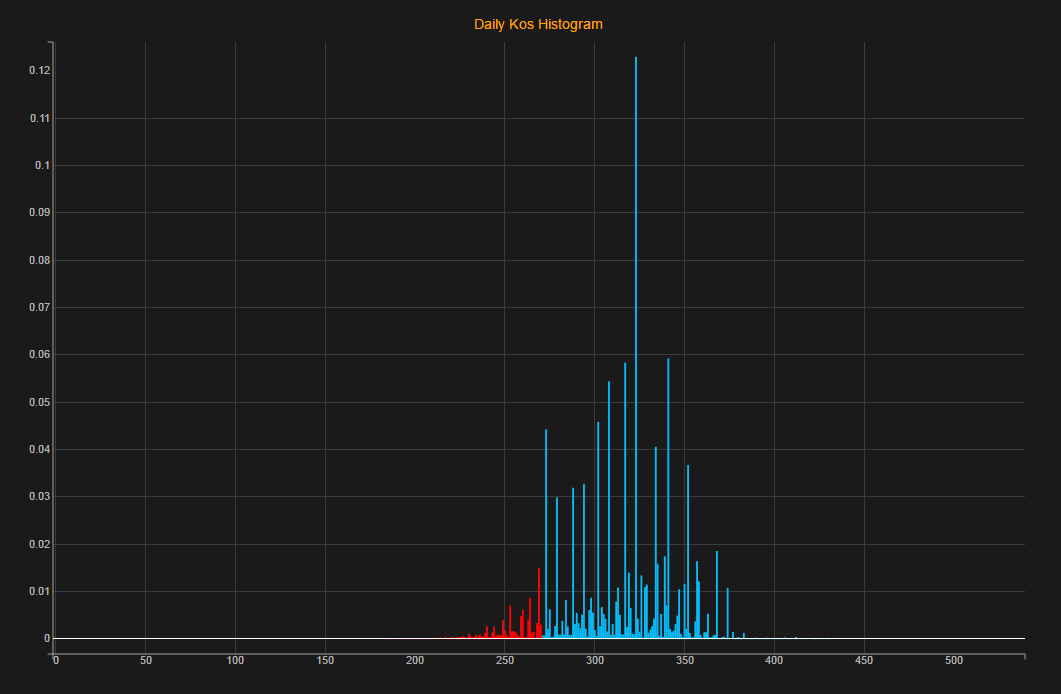}
\caption{The DK and PEC histogram are very similar in terms of variance, but the methodology of DK lends itself to spikes at different EV's. The Selten unduly penalizes this as there is no topology induced amongst the different bins.}
\label{dk-histogram}
\end{center}
\end{figure}

\begin{figure}[h]
\begin{center}
\includegraphics[width=16.5cm, height=7.2cm]{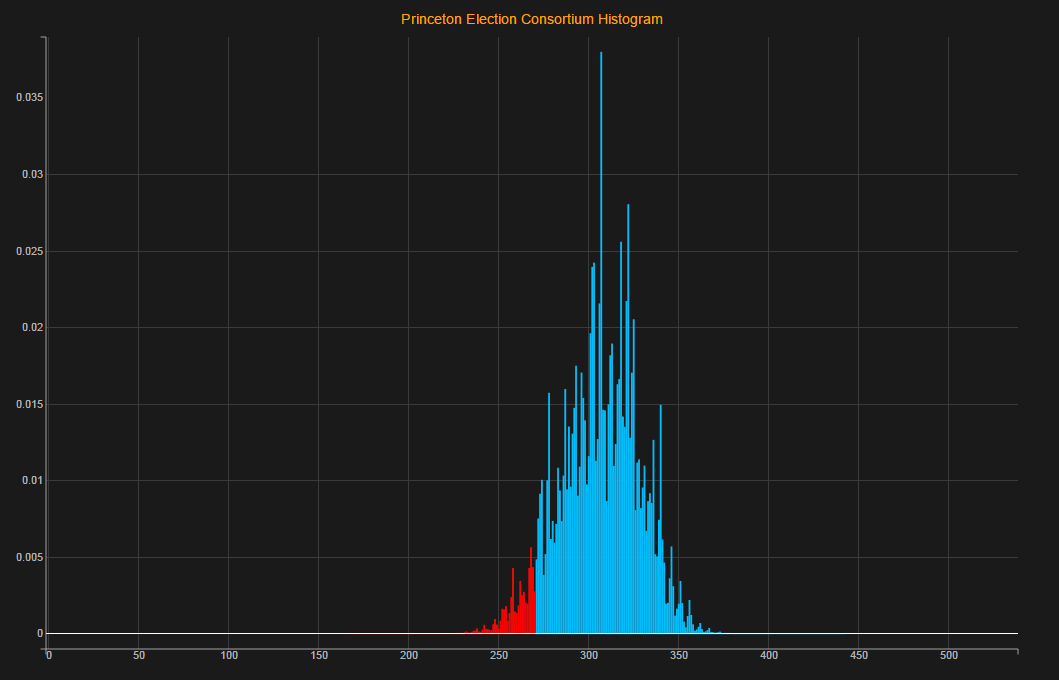}
\caption{The DK and PEC histogram are very similar in terms of variance, but the methodology of DK lends itself to spikes at different EV's. The Selten unduly penalizes this as there is no topology induced amongst the different bins. Figure \ref{overall-cdf} shows that they are actually very similar.}
\label{pec-hist}
\end{center}
\end{figure}

\begin{figure}[h]
\begin{center}
\includegraphics[width=16.5cm, height=7.cm]{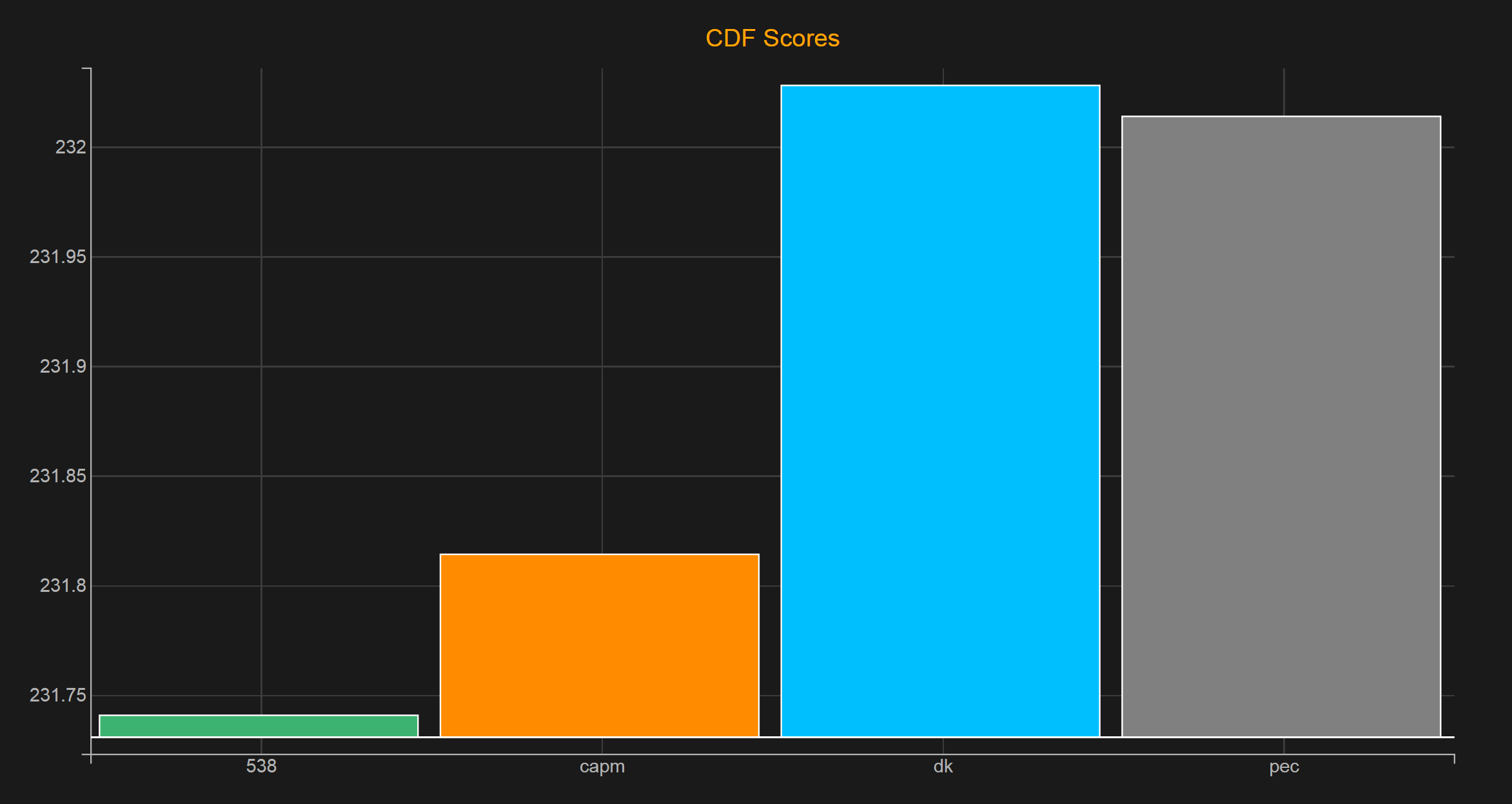}
\caption{The flaws of the Selten are corrected by the CDF Score, which shows that DK and PEC were very close in terms of a distributional score.}
\label{overall-cdf}
\end{center}
\end{figure}

\begin{figure}[h]
\begin{center}
\includegraphics[width=16.5cm, height=7.2cm]{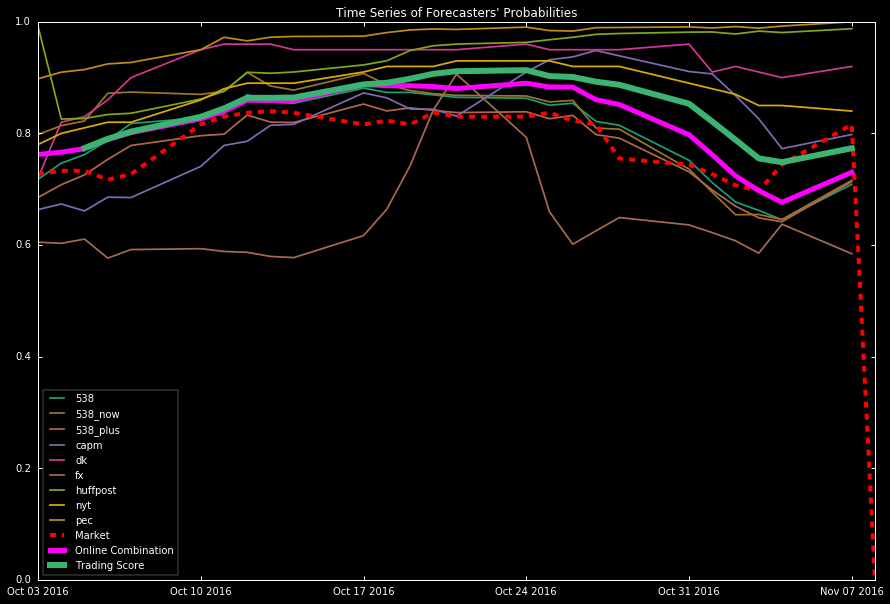}
\caption{Time Series of Modellers, we see that the online learning Algorithm (magenta) picks up the drop in Clinton quite quickly.}
\label{model-ts}
\end{center}
\end{figure}

\begin{figure}[h]
\begin{center}
\includegraphics[width=16.5cm, height=7.cm]{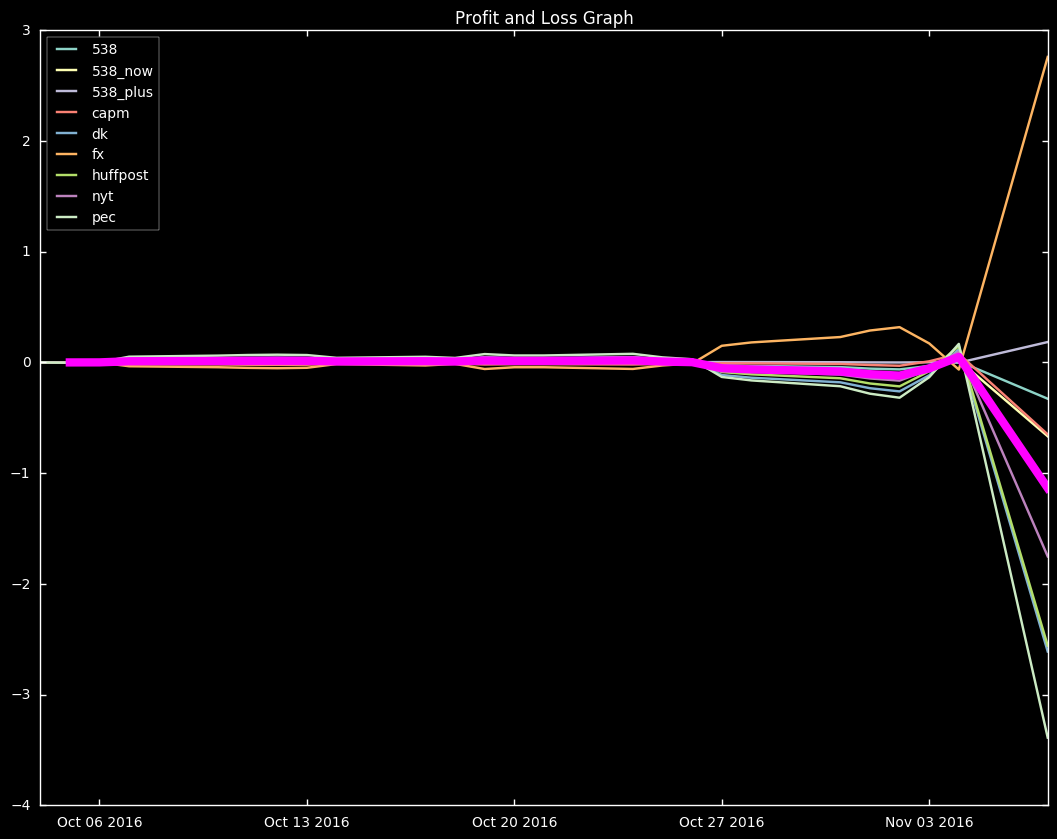}
\caption{Despite being net long, the online learning algorithm (magenta) does not end up with the worst loss.}
\label{pl-ts}
\end{center}
\end{figure}

\begin{figure}[h]
\begin{center}
\includegraphics[width=16.5cm, height=7.cm]{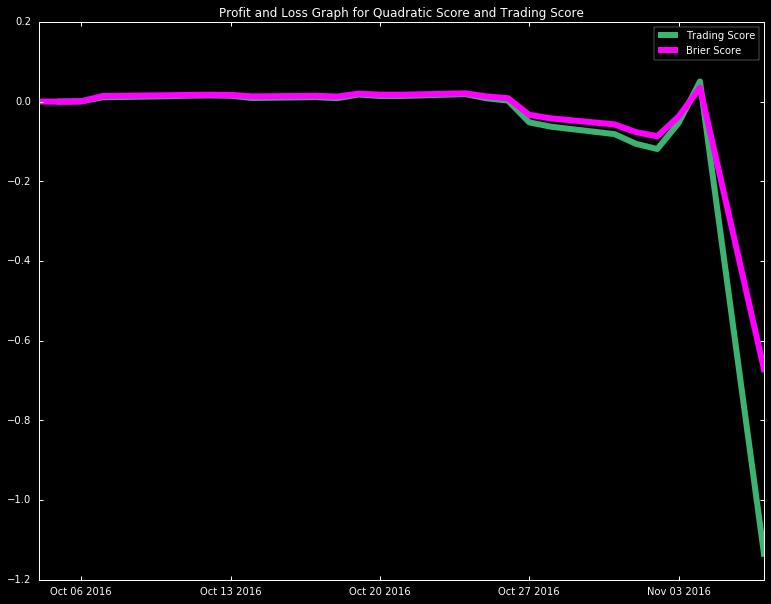}
\caption{As we see, the Quadratic Score has a higher profit and loss.}
\label{pl-ts-quad}
\end{center}
\end{figure}

\begin{figure}[h]
\begin{center}
\includegraphics[width=16.5cm, height=7.cm]{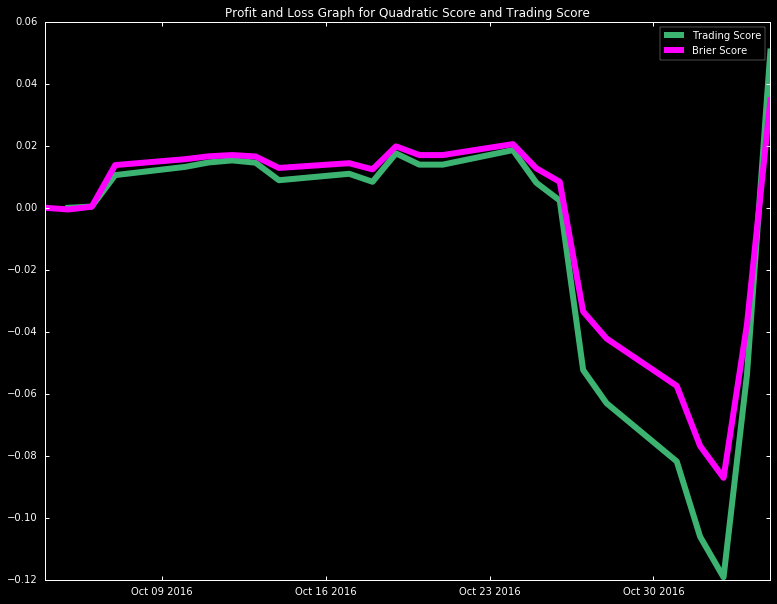}
\caption{However, if we exclude the final result, trading with the trading score would have generated a higher P\&L.}
\label{pl-nolast}
\end{center}
\end{figure}

\begin{figure}[h]
\begin{center}
\includegraphics[width=16.5cm, height=7.cm]{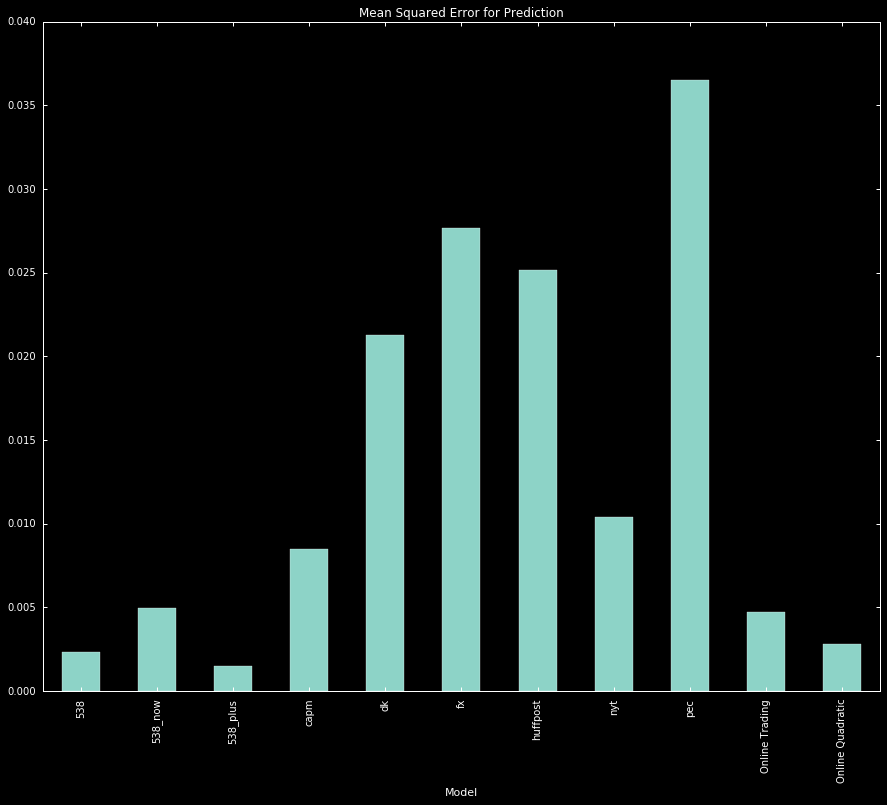}
\caption{As we see, the Online Quadratic Model and the Online Trading Score had similar performance losing only to 538 in predicting the betting market.}
\label{mse}
\end{center}
\end{figure}

\newpage

\section{Behaviour of Different Scoring Functions}

\begin{figure}[h]
\begin{center}
\includegraphics[width=16.5cm, height=15cm]{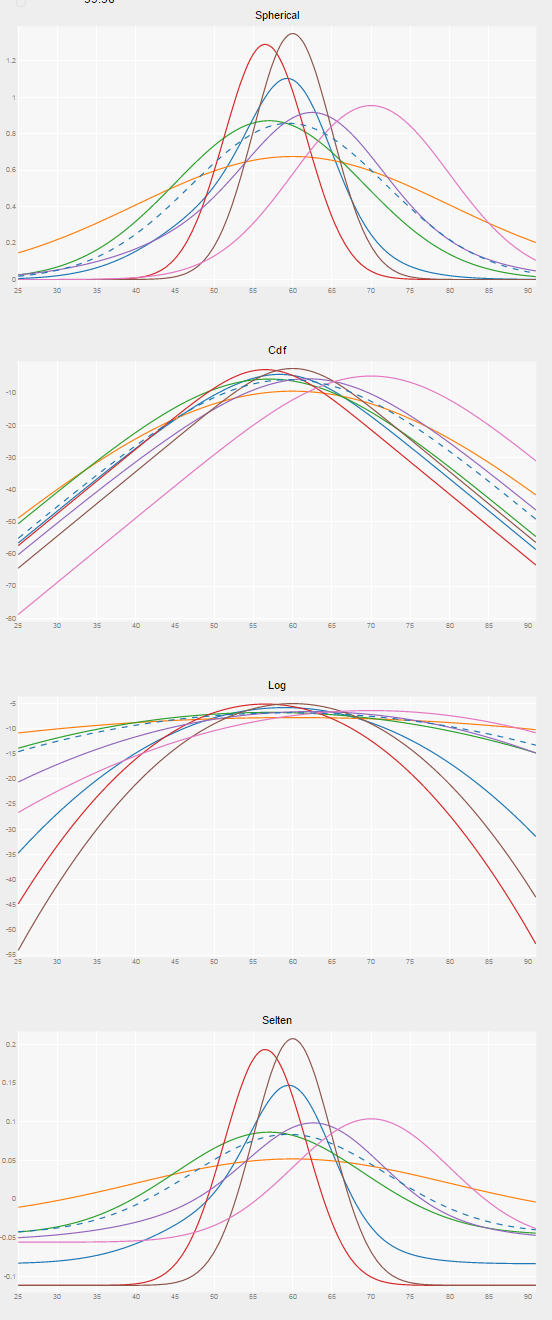}
\caption{Plot of the different scoring functions with the x-axis representing the realization and the y-axis representing the Score. The distributions used are Gaussians with different means and standard deviations. As we see, the Selten and Spherical are flat at the tail, while the log goes parabolically to $-\infty$, while the CDF score goes linearly to $-\infty$.}
\label{shape-fig}
\label{score-comp}
\end{center}
\end{figure}

\newpage

\section{Simulation Results}

\begin{figure}[h]
\begin{center}
\includegraphics[width=16.5cm, height=10cm]{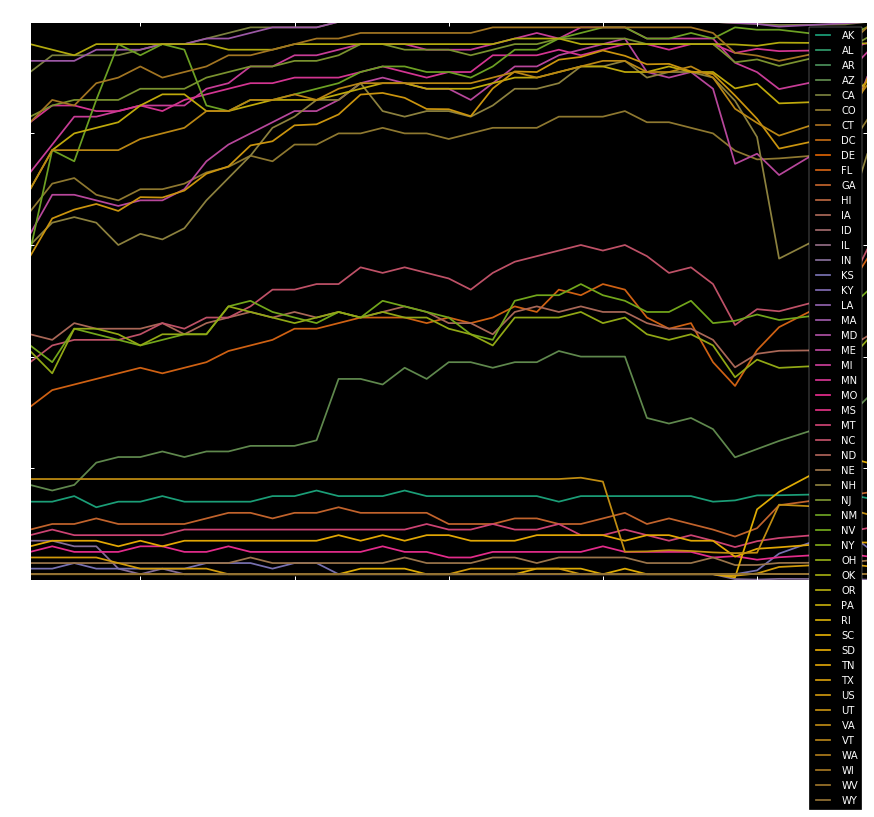}
\caption{Each line shows the time series of probabilities for each state.}
\label{capm-ts}
\end{center}
\end{figure}

\begin{figure}[h]

\begin{center}
\includegraphics[width=16.5cm, height=10cm]{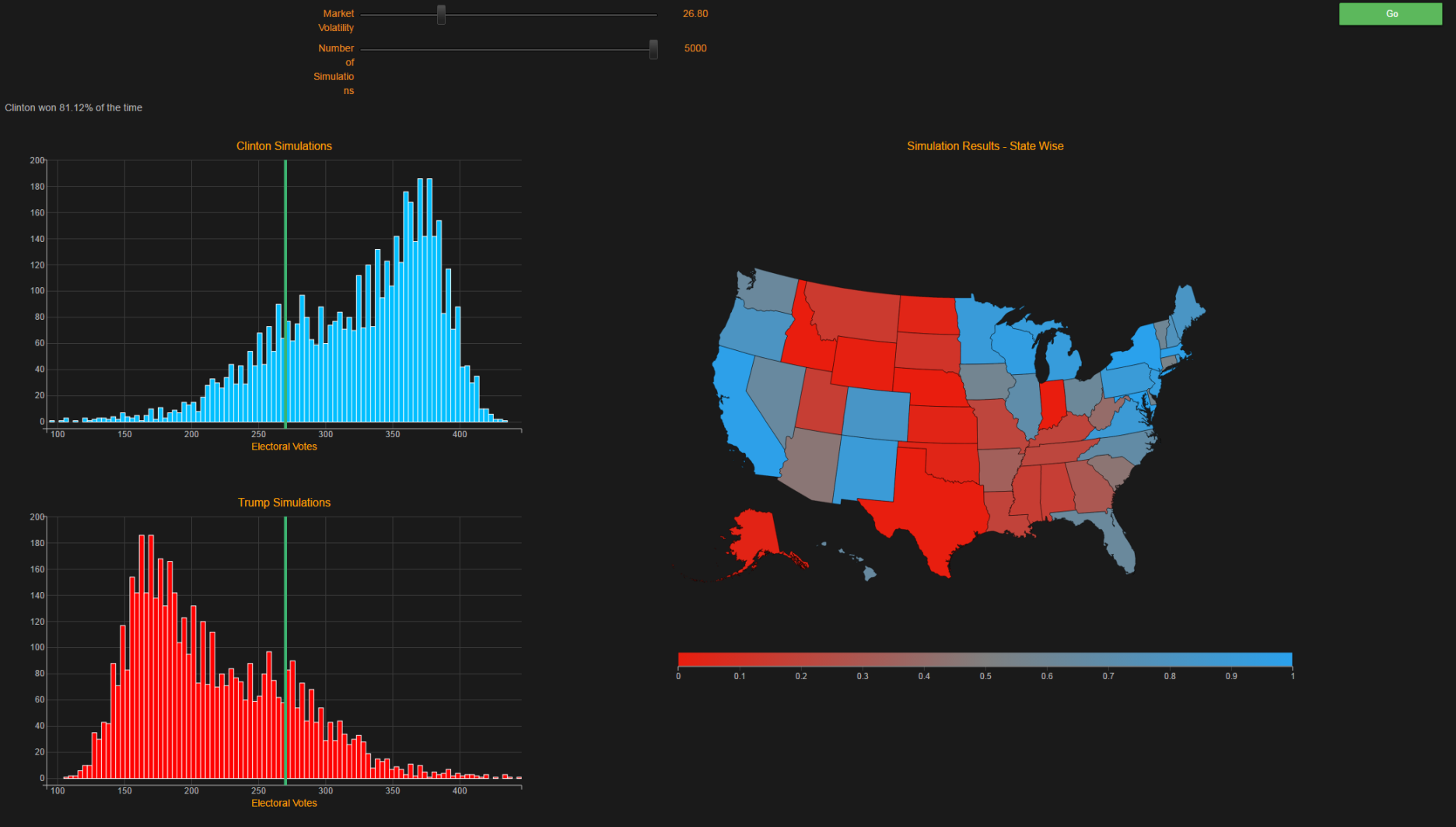}
\caption{Plot of the map for representing the probabilities for each state, red implies republican and blue democrat. The histogram has the number of simulations on the y axis and the Electoral Votes won on the X-axis.}
\label{capm-sim}
\end{center}
\end{figure}

\begin{figure}[h]
\begin{center}
\includegraphics[width=16.5cm, height=10cm]{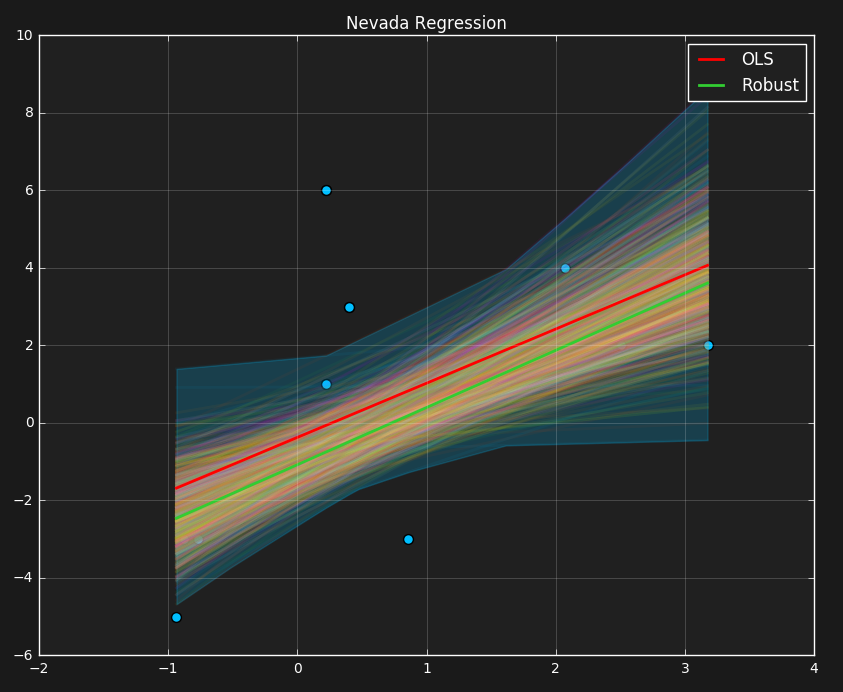}
\caption{Here we see the samples drawn from the predictive distribution of the $\alpha$'s and $\beta$'s. We use each one to generate a noise about the state.}
\label{nevada-ts}
\end{center}
\end{figure}

\begin{figure}[h]
\begin{center}
\includegraphics[width=16.5cm, height=10cm]{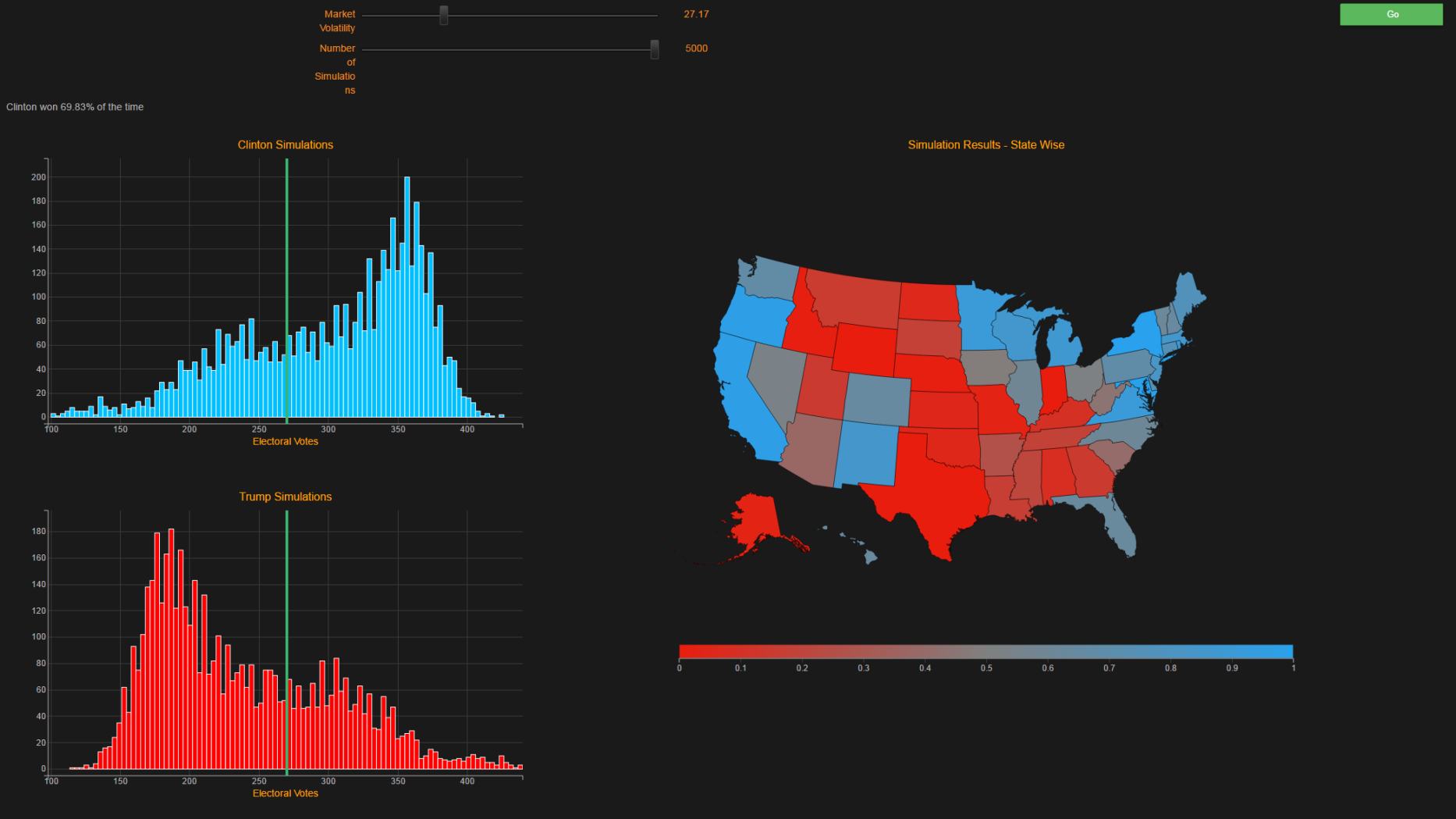}
\caption{Plot of the map for representing the probabilities for each state, red implies republican and blue democrat. The histogram has the number of simulations on the y axis and the Electoral Votes won on the X-axis. As we see the Student T distribution gives a very similar probability with fatter tails for Trump.}
\label{student-sim}
\end{center}
\end{figure}

\end{document}